\documentclass[prl,twocolumn,amssymb,floatfix,superscriptaddress]{revtex4-2}

\usepackage{ulem}
\usepackage{graphicx}
\usepackage{amsmath}
\usepackage{amstext}
\usepackage{amssymb}
\usepackage[colorlinks,citecolor=blue]{hyperref}
\usepackage{graphicx}
\usepackage{tabularx}
\usepackage{array}
\usepackage{amsmath}
\usepackage{amstext}
\usepackage{longtable}
\usepackage{amssymb}
\usepackage{amsfonts}
\usepackage{longtable,booktabs}
\usepackage{hyperref}
\usepackage{url}
\usepackage{dsfont,bbm}
\usepackage[usenames,dvipsnames]{color}
\usepackage{amsbsy}
\usepackage{dcolumn}
\usepackage{amsthm}
\usepackage{bm}
\usepackage{tikz}
\usepackage{multirow}
\usepackage{hyperref}
\hypersetup{
    colorlinks=true,
    linkcolor=black,
    filecolor=black,
    urlcolor=black,
}

\usepackage{mathrsfs}
\usepackage{amsfonts}
\usepackage{amsbsy}
\usepackage{dcolumn}
\usepackage{bm}
\usepackage{multirow}
\usepackage{color}

\newcommand{\da}{\downarrow}

\newcommand{\bk}{\mathbf k}

\renewcommand{\vec}[1]{\mathbf{#1}}
\newcommand{\vk}{{\vec{k}}}

\begin{document}
\title{Density matrix renormalization group study of quantum-geometry-facilitated pair density wave order}

	\author{Hao-Xin Wang}
           \email{haoxinwang@cuhk.edu.hk}
	\address{Department of Physics, The Chinese University of Hong Kong, Sha Tin, New Territories, Hong Kong, China}
	\author{Wen Huang}
	\email{huangw3@sustech.edu.cn}
	\address{Shenzhen Institute for Quantum Science and Engineering, Southern University of Science and Technology, Shenzhen 518055, Guangdong, China}
	\address{International Quantum Academy, Shenzhen 518048, China}
	\address{Guangdong Provincial Key Laboratory of Quantum Science and Engineering, Southern University of Science and Technology, Shenzhen 518055, China}

\date{\today}
\begin{abstract}
Understanding the formation of novel pair density waves (PDWs) in strongly correlated electronic systems remains challenging. 
Recent mean-field studies suggest that PDW phases may arise in strong-coupling multiband superconductors by virtue of the quantum geometric properties of paired electrons. However, scrutiny through sophisticated many-body calculations has been lacking. Employing large-scale density matrix renormalization group calculations, we obtain in the strong-coupling regime the phase diagram as a function of doping concentration and a tuning interaction parameter for a simple two-orbital model that incorporates quantum geometric effects. The phase diagram reveals a robust PDW phase spanning a broad range of parameters, characterized by a Luttinger parameter $K_{sc} \sim 0.3$ and the absence of coexisting competing spin or charge density wave  orders. The observed pairing field configuration aligns with the phenomenological understanding that quantum geometry can promote PDW formation. Our study provides the most compelling numerical evidence to date for quantum-geometry-facilitated intrinsic PDW order in strongly correlated systems, paving the way for further exploration of novel PDW orders and quantum geometric effects in such systems.
\end{abstract}

\maketitle

{\bf Introduction.--} Pair density waves (PDWs) are superconducting states wherein the superconducting order parameter exhibits periodic modulation. A classic example of such a novel state of matter is the Fulde-Ferrell-Larkin-Ovchinnikov state~\cite{Fulde:64,Larkin:65}, which may appear in BCS superconductors subject to external magnetic field. A more difficult and modern puzzle is the formation of PDW order in strongly correlated electronic systems~\cite{Fradkin:15, Agterberg:20}, including in cuprate and iron-based superconductors~\cite{Hamidian:16,Ruan:18,Edkins:19,DuZ:20,LiX:21, lee2023pairdensity, Liu2023PDW_Nature,ZhangYaoArXiv2024}. Since understanding the microscopic mechanism driving the PDW may hold implications to the understanding of high-$T_c$ superconductivity (SC), many theoretical studies have emerged in this direction~\cite{Berg2010, Jaefari2012, Han2020StrongPRL, wang2024pair, Shaffer2023Triplet, Shaffer2023Weak, Wu2023PairPRB, Castro2023Emergence, liu2024enhanced, ticea2024pair, Chen2023Singlet, Tsvelik2023, Wu2023Sublattice, Han2022Pair, yue2024pseudogap, Donna:23dwave, Patrick2014Amperean, Wu2023Pair, Haokai2023, Setty2023Exact, Jiang2023Pair, JiangAndTom2023Pair-Frontiers,Jin2022PRL,Banerjee2022Charge,Song2022Doping,jiangyifan2023pair,GuoHM2024,Kivelson2020NPJQuantumM}. However, a full comprehension is still lacking~\cite{Fradkin:15, Agterberg:20}.  

\begin{figure}[tbp]
    \centering
    \includegraphics[width=0.48\textwidth]{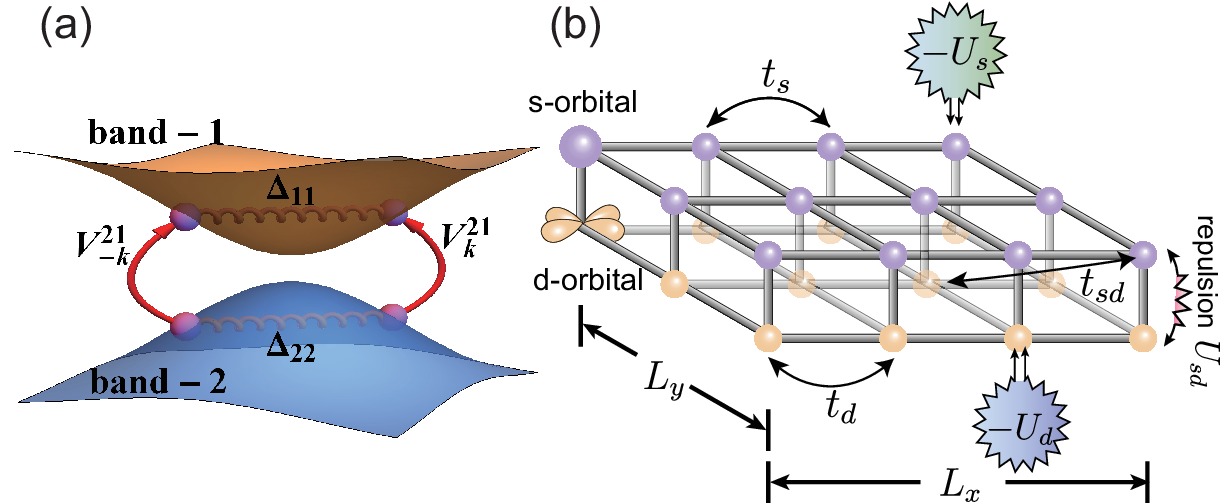}
    \caption{(a) A representative manifestation of quantum geometric effects in multiband superconductors. Thanks to the interband velocity, Cooper pairs can tunnel between different Bloch bands, resulting in effective Josephson coupling between their order parameters. (b) Sketch of bilayer lattice model used in our DMRG simulation, with $s$ and $d_{xy}$ orbitals residing on separate layers. Details are provided in the text.}
    \label{fig:QGsketch}
\end{figure}

Recently, it was proposed that the quantum geometry of paired Bloch electrons in strongly-coupled multiband superconductors may provide an intrinsic mechanism for PDW~\cite{ChenW:23, JiangG:23, ZhaoyuHan2024GQN}. The quantum geometry of a Bloch state is formally described by the quantum geometric tensor~\cite{Provost1980Riemannian,Peotta:15,Ahn2022NaturePhysics}, whose real and imaginary parts define respectively the quantum metric and Berry curvature. Quantum geometry impacts the transport and electromagnetic responses of multiband systems, manifesting most famously in the context of topological band theories~\cite{Thouless:82,Qi:11,Xiao:10,Hasan:10} --- where the Hall response is essentially a Berry curvature effect. Recent years have witnessed the study of quantum geometric effects reaching a much broader spectrum of condensed matter physics~\cite{Torma:23,LiuTianyuNSR2024}, including multiband SC~\cite{Peotta:15,Julku:16, Liang:17,Torma:22,Iskin2018,Iskin2019,Wangzhiqiang2020,VermaPNAS2021,Chen:21,AhnJ2021,YanaseAnapole2023,YanaseDiode2023,ChenS:24,ZhangJL2024, Debanjan2023PRLSuperconductivity,DanMao2023PNAS,HuYJ2024}. In particular, the superfluid weight, which measures the ability of Cooper pairs to sustain macroscopic phase coherence, was found to contain a contribution closely related to the geometric properties of the paired electrons~\cite{Peotta:15,Julku:16,Liang:17,HuYJ2024}. This geometric superfluid weight has been invoked to explain the flatband SC in several model systems~\cite{Peotta:15,Julku:16}, including in twisted bilayer graphene~\cite{Hu:19,Julku:20,Xie:20,TianH:23,LawKT2023}, where traditional theory would have otherwise concluded that superfluidity is unsustainable as electrons in flatbands are immobile.

Lately, based on the observation that the geometric superfluid weight can be negative under certain circumstances, two independent mean-field studies proposed that quantum geometry has the potential to facilitate the formation of PDW order~\cite{ChenW:23,JiangG:23}. However, in these studies, PDW emerges at rather strong coupling, for which many-body effects such as the loss of quasiparticle coherence and the potential mutual influence of multiple ordering tendencies cannot be properly accounted for within mean-field. Hence, these results are questionable and confirmation of this geometric mechanism using sophisticated many-body approaches is highly desired.

In this study, we perform large-scale density matrix renormalization group (DMRG)~\cite{White1992PRL, White1993PRB} simulations of a simple strongly interacting two-band model devised in Ref.~\onlinecite{ChenW:23}. Our numerical results demonstrate that the PDW phase can stabilize in the strong coupling regime, supporting the scenario in which quantum geometry plays a pivotal role in stabilizing this exotic order.


{\bf Model and method.---} The two-orbital model consists of an $s$- and a $d_{xy}$-orbital on a square lattice. The non-interacting part of the Hamiltonian, if written in momentum-space, is given by $H_0 = \sum_{\vk,\sigma} H_{0,\vk}$ with,
\begin{equation}
H_{0,\vk}=\xi_{s\vk} c^\dagger_{s\vk}c_{s\vk}+ \xi_{d\vk} c^\dagger_{d\vk}c_{d\vk} + \lambda_\vk (c^\dagger_{s\vk}c_{d\vk}+c^\dagger_{d\vk}c_{s\vk}). 
\label{eq:H0}
\end{equation}
Here, the dispersion relation $\xi_{\alpha\vk}=-2t_\alpha(\cos k_x +\cos k_y)$ ($\alpha=s,d$  the orbital index) and $\lambda_\vk=4t_{sd}\sin k_x \sin k_y$, where $t_{s(d)}$ denotes the $s (d)$-orbital nearest-neighbor intraorbital hopping and $t_{sd}$ the next-nearest-neighbor interorbital mixing. Direct nearest-neighbor hopping between the $s$- and $d_{xy}$-orbitals is forbidden by symmetry. 
For brevity, we omit the spin indices in Eq.~\eqref{eq:H0}, and henceforth refer to the $d_{xy}$-orbital as the $d$-orbital. 
Diagonalizing this Hamiltonian results in two bands, whose wavefunctions encode the geometric information crucial to the formation of PDW. Note that, this model is topologically trivial, in the regard that the Berry curvature of the Bloch bands vanishes. In practice, we keep a balance between $t_{sd}$ and $t_s-t_d$ to ensure a sizable quantum geometric effect. Without loss of generality, we set $t_s=-t_d=t_{sd}=1$ throughout the study. 

The interacting part of the Hamiltonian is described by $H_\text{int} = \sum_i H_{\text{int},i}$ where, on each site $i$, 
\begin{eqnarray}
H_{\text{int},i} &=& -U_{s} c^\dagger_{i,s\uparrow}c^\dagger_{i,s\downarrow} c_{i,s\da}c_{i,s\uparrow} -U_{d} c^\dagger_{i,d\uparrow}c^\dagger_{i,d\da} c_{i,d\da}c_{i,d\uparrow} \nonumber \\
&& + U_{sd} \left[ c^\dagger_{i,s\uparrow}c^\dagger_{i,s\da} c_{i,d\da}c_{i,d\uparrow}  + (s \leftrightarrow d)\right] \,.
\label{eq:interactions}
\end{eqnarray}
Here, $-U_{s}, -U_{d}$ designate the attractive interactions that drive onsite intraorbital spin-singlet pairings $\Delta_s$ and $\Delta_d$, and $U_{sd}$ labels the repulsive interorbital pair hopping interaction. For a superconducting state developed at strong $U_{sd}$, $\Delta_s$ and $\Delta_d$ tend to acquire a phase difference of $\pi$, referred to as the $\hat{\Delta}_{+-}$ configuration. The alternative scenario with no phase difference is referred to as the $\hat{\Delta}_{++}$ configuration.

\begin{figure}[tbp]
    \centering
    \includegraphics[width = 0.48\textwidth]{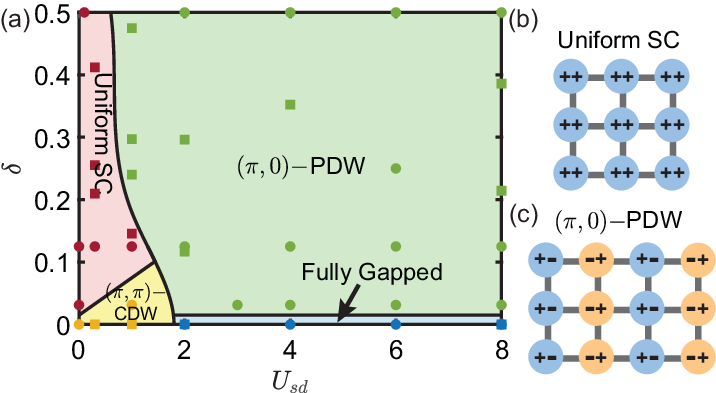}
    \caption{ (a) The quantum phase diagram as a function of doping level $\delta$ and the inter-orbital repulsive pair hopping $U_{sd}$, obtained from $L_y = 4$ DMRG calculations. The intra-orbital interaction strength is fixed at $U_{s}=U_{d}=8$. Circles indicate calculations performed using canonical ensemble DMRG, while squares represent those gathered using grand canonical ensemble DMRG. (b) and (c), illustration of the uniform SC and $(\pi,0)$-PDW states obtained in our calculations. The symbol '$++$' in (b) indicates $\hat\Delta_{++}$ configuration on each lattice site; while the symbols '$+-$' and '$-+$' in (c) denote $\hat\Delta_{+-}$ configuration on each site, with an additional overall phase modulation in $x$-direction. 
}
    \label{fig: phase diagram}
\end{figure}

In our real-space DMRG calculation, we map the Hamiltonian \eqref{eq:H0} and \eqref{eq:interactions} into a bilayer square lattice model, where each layer represents one orbital degree of freedom [see Fig.~\ref{fig:QGsketch} (b)]. The calculations are performed on a $L_x \times L_y$ lattice with open boundary conditions along the $x$-direction and periodic boundary conditions along the $y$-direction. The total number of sites is thus $N = 2L_x L_y$. Data presented in the maintext are all for $L_y = 4$, while we also present some $L_y = 3$ calculations in Supplementary Materials~\cite{Supplementary} to check consistency. Simulating such a bilayer $L_y = 4$ ladder is computationally roughly as demanding as the most state-of-the-art simulation of the single-orbital Hubbard and $t$-$J$ models with 8-leg ladder~\cite{HongChen2019Science, Shiwei2020Absence, Shoushu2021Robust, Hongchen2021HighTemperature, Donna:23dwave, YiFan2024Ground, Donna2024Emergent, SWZhang2024Coexistence}. Throughout the maintext, we focus on the strong coupling regime and set $U_s = U_d  = 8$, which is computationally more accessible to us. We tune the value of $U_{sd}$ and the doping level $\delta = 1 - \sum_{i, \alpha = s,d} n_{i, \alpha} /N  > 0$ with $n_{i,\alpha}$ denoting the occupancy of the $\alpha$-orbital on site $i$, to explore different phases of the model. Thanks to the particle-hole symmetry, we only need to study the hole-doped regime.  

In this study, we utilize two types of DMRG techniques, one imposing $U(1)_{\mathrm{charge}} \otimes U(1)_{\mathrm{spin}}$ symmetry and the other solely enforcing $U(1)_{\mathrm{spin}}$ symmetry, and they serve as cross-checks for each other. 
The former is hereby referred to as the canonical ensemble calculation.
The latter, termed the grand canonical ensemble, allows for a non-zero estimate of the order parameter magnitude in the thermodynamic limit \cite{Shengtao:21Ground, Shengtao:22Pairing, Shengtao:23Density, Donna:23dwave}. In the former, we retain the bond dimension up to $D = 24000$, ensuring a truncation error $\varepsilon \sim 2 \times 10^{-6}$. We also perform various simulations with different initial states to eliminate metastable states. All data from the canonical ensemble calculations presented below have been extrapolated to the $\varepsilon \rightarrow 0$ limit using quadratic fitting \cite{Supplementary}.

{\bf PDW order. ---}  Figure \ref{fig: phase diagram} presents a representative phase diagram as a function of doping concentration $\delta$ and pair hopping interaction $U_{sd}$. The diagram reveals a PDW phase dominating in a wide range of doping and pair hopping interaction, while a uniform superconducting phase develops at finite doping and small $U_{sd}$. Around zero doping, a charge density wave (CDW) with momentum $(\pi,\pi)$ initially stabilizes at weak $U_{sd}$, while a fully gapped featureless phase appear at stronger $U_{sd}$.

\begin{figure}[b]
    \centering
    \includegraphics[width=0.48\textwidth]{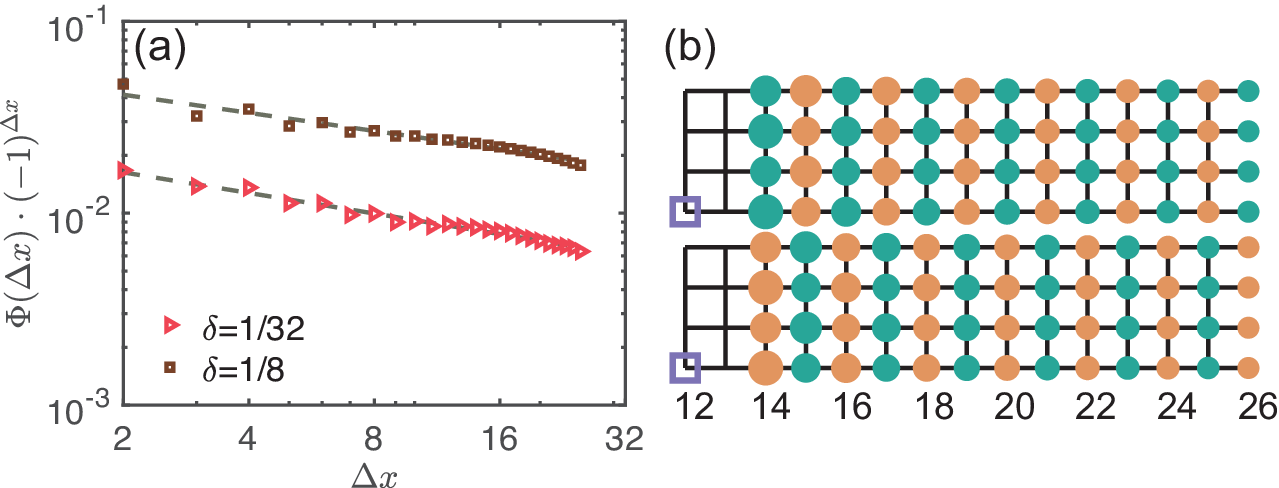}
    \caption{SC Correlation functions in PDW phase. (a) SC correlations on the $s$-orbital along the horizontal direction and their power-law fitting. 
    (b) SC correlations on the 4-leg lattice with doping $\delta = 1/8$. The reference is the $s$-orbital intra-orbital pairing on site ${\bf r_0} = (12, 1)$. The upper panel shows the intra-$s$-orbital correlation, and the lower panel displays the inter-orbital correlation between $s$- and $d$-orbitals. The area of the colored dots encode the magnitude of the correlation, and their color indicate the sign of the correlation, {\it i.e.}, green (orange) is positive (negative).}
    \label{fig: PDW Corr}
\end{figure}

We first concentrate on the PDW phase, the central result of this study. SC is described by the equal-time pair-pair correlation function defined as
\begin{equation}
    \Phi_{\alpha\beta}({\bf r} = {\bf r_i} - {\bf r_j}) = \langle \mathcal{O}_{i, \alpha}^\dagger \mathcal{O}_{j, \beta}\rangle,
    \label{eq:pairCorr}
\end{equation}
where $\mathcal{O}_{i, \alpha}^\dagger = c^\dagger_{i, \alpha,\uparrow} c^\dagger_{i, \alpha, \downarrow}$ is the spin-singlet pair creation operator on site $i$ with orbital index $\alpha$. To reduce boundary effects, the reference point ${\bf r_j}$ is fixed at ${\bf r_j} = (x_j, y_j) = (L_x/4, 1)$ on the $s$-orbital. Figure \ref{fig: PDW Corr} (a) shows the pair-pair correlation function along the horizontal direction $\Phi(\Delta x) := \Phi_{ss}(\Delta x, \Delta y=0)$ for $U_{sd} = 8$ and $\delta = 1/8$ and $1/32$ within the PDW phase. This correlation exhibits characteristic power-law decay $\Phi(r) \propto r^{-K_{sc}}$ with Luttinger parameters $K_{sc} \simeq 0.36$ and $0.31$, respectively. The power-law decay with $K_{sc} < 2$ indicates the presence of quasi-long-range superconducting order with a diverging superconducting susceptibility. Moreover, as shown in the Supplementary Materials \cite{Supplementary}, $K_{sc}$ decreases as $L_y$ increases from 3 to 4. This suggests stronger pairing as the ladder widens, further consolidating the long-range SC order in the two-dimensional limit. 

Figure \ref{fig: PDW Corr} (b) illustrates the correlation function $\Phi_{\alpha\beta}({\bf r})$ across the lattice. The area and color of filled circles encode the magnitude and sign of the correlation, respectively. The correlation alternates in sign with a periodicity of two lattice constants along the $x$-direction and is translational invariant along $y$. The PDW order, therefore, has a modulation wavevector $(\pi,0)$. Notably, at each location ${\bf r_i}$, the signs of the intraorbital pairings on the two orbitals are opposite, corresponding to the $\hat{\Delta}_{+-}$ configuration. In our calculations, the PDW phase always appears in this configuration.

\begin{figure}
    \centering
    \includegraphics[width=0.48\textwidth]{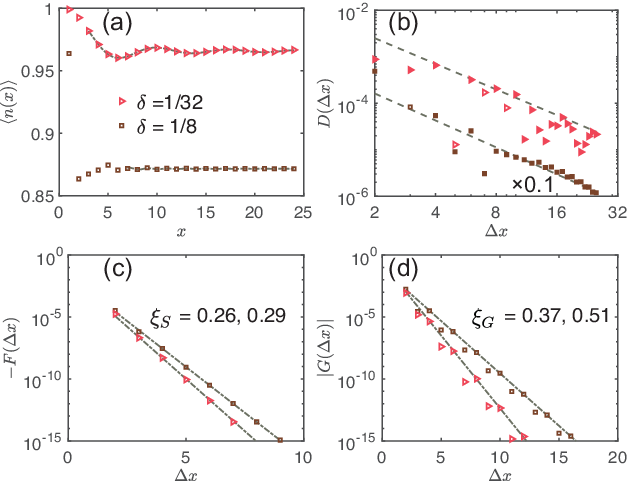}
    \caption{Charge density and different correlation functions in the PDW phase with doping $\delta=1/32$ (triangles) and $\delta=1/8$ (squares).
    (a) Charge density profile of the $s$-orbital in half of the system. The other half is related by a reflection.
    (b) Intra-$s$-orbital density-density correlation functions with power-law fitting. Filled points indicate negative values while empty points indicate positive values.
    (c) and (d) display the intra-$s$-orbital spin correlation functions and single-particle correlation functions, respectively. }
    \label{fig: PDW Corr2}
\end{figure}

To check if there are coexisting or competing instabilities, we also examine the possibility of charge and spin ordering. We first evaluate the charge density distribution $n_\alpha(x)  = \frac{1}{L_y} \sum_y \langle \hat n_\alpha(x,y) \rangle$. Similar to the PDW order, the charge density does not exhibit any modulation along the $y$ direction. The density profile along $x$ direction associated with the two calculations in Fig.~\ref{fig: PDW Corr} are shown in Fig.~\ref{fig: PDW Corr2} (a). 
Under the chosen parameter set $t_s = -t_d$ and $U_s = U_d$, the two orbitals have the same density. At doping $\delta = 1/32$, on top of a large uniform component, one can see a clear oscillatory component whose amplitude decays rapidly away from the open boundary. At doping $\delta = 1/8$, the oscillation is almost invisible in the bulk. The charge density modulations can be well described by Friedel oscillations,
\begin{equation}
    n_\alpha(x)=A\cos (Q \cdot r+\phi) x^{-K_{c} / 2}+n_0
\end{equation}
where $n_0$ is the average density, $Q$ is the charge modulation wavevector, $A$ and $\phi$ are some non-universal constants, and $K_c$ is the Luttinger parameter. By fitting the density profiles for $\delta = 1/32$, we obtain $Q\approx \frac{2\pi}{8}$ and $K_{c} \approx 2.94$. The value of $Q$ indicates a scenario of $\frac{1}{L_y}$-filled stripy charge density modulation. These results provide preliminary evidence for the absence of CDW instability in this regime of the phase diagram. To substantiate this conclusion, we further measure the density correlation function
\begin{equation}
    \begin{aligned}
     D_{\alpha, \beta}(r)  =& \frac{1}{L_y} \sum_y \langle \hat n_\alpha(x_0,y) \hat n_\beta(x_0 + r, y)  \rangle \\
     &- \langle \hat n_\alpha(x_0,y) \rangle \langle  \hat n_\beta(x_0+r,y)\rangle .
\end{aligned}
\end{equation}
For both orbitals (see Fig.~\ref{fig: PDW Corr2} (b) and Supplemental Materials~\cite{Supplementary}), we see power-law decay form $D_{\alpha \beta}(r) \sim r^{-K^{\alpha\beta}_c}$, with Luttinger parameter $K^{ss}_{c}\simeq 1.83,~1.95$ for the two respective dopings. The power-law decay density correlation, combined with the Friedel oscillations, suggest that the PDW phase is charge gapless and that CDW order, if any, must be significantly weaker than PDW. 


Further, the calculated spin correlation function $F_{\alpha,\beta}(r) = \frac{1}{L_y} \sum_{y=1}^{L_y} \langle \mathbf{S}_\alpha(x_0, y) \cdot \mathbf{S}_\beta(x_0 + r, y)\rangle$, and the single-particle correlation function $G_{\alpha,\beta}(r) =  \frac{1}{L_y} \sum_{y=1}^{L_y} \langle c_{\alpha}^\dagger(x_0, y) c_{\beta} (x_0 + r, y) \rangle$, as shown in Fig.~\ref{fig: PDW Corr2} (c) and (d), both decay exponentially fast with respective correlation lengths $\xi_S$ and $\xi_G$ less than a lattice constant. These indicate large spin and single-particle gap, and that spin order is absent in the PDW phase.

Combining the above correlation functions, the PDW states qualitatively aligns with the expectation for Luther-Emery liquids \cite{Giamarchi_book}. Additionally, we present results of the PDW phase obtained for $L_y = 3$ and provide grand canonical ensemble calculations in the Supplementary Material~\cite{Supplementary}. A comprehensive analysis indicates that the PDW instability is the only one that is enhanced as the system width $L_y$ increases.

\begin{figure}[htbp]
    \centering
    \includegraphics[width=0.49\textwidth]{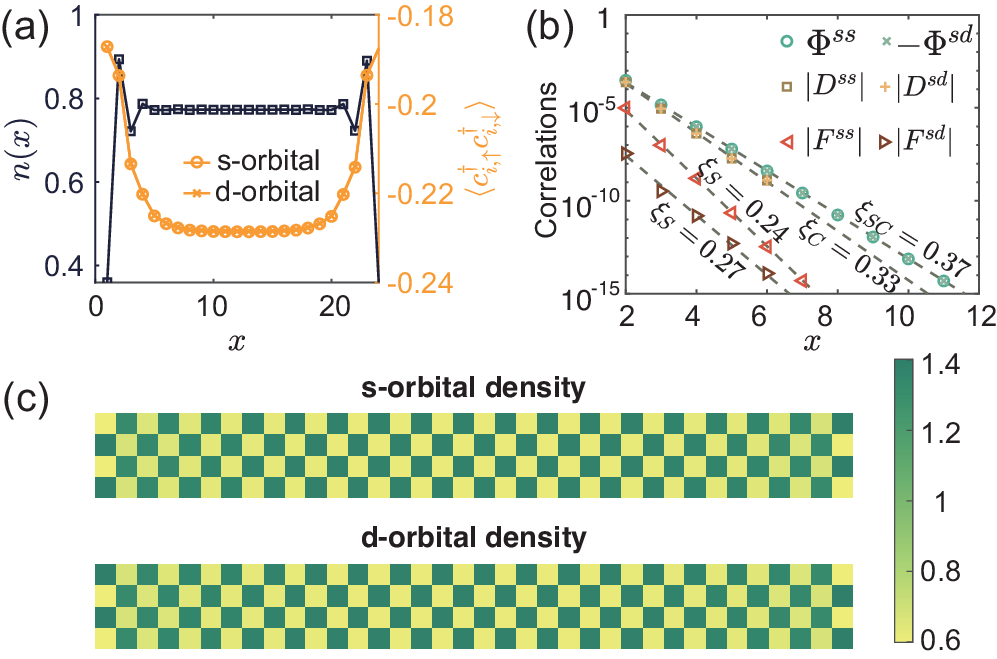}
    \caption{(a) Charge denisty and SC pair field in the uniform SC phase for $U_{sd}=0.3$, $\mu_s = \mu_d = -4.7$, corresponding to approximate doping level $\delta = 0.25$. (b) Correlation functions at half filling and $U_{sd}=8$. (c) Charge density pattern at half filling, $U_{sd} = 0$, showing a CDW with modulation wavevector $(\pi,\pi)$. The data are obtained from grand canonical ensemble DMRG with (a) $D=10000$ and (b, c) $D=20000$ states. }
    \label{fig: other phases}
\end{figure}

{\bf Uniform SC, CDW and featureless phases ---}
Having established and characterized the PDW phase, we now turn to the remaining phases in the phase diagram of Fig.~\ref{fig: phase diagram}. In Fig.~\ref{fig: other phases} (a), we provide the charge density profile and the pairing potential $\langle c_{\alpha,i,\uparrow} c_{\alpha,i,\downarrow}\rangle$, for a representative set of parameters in the uniform superconducting phase,  calculated using grand canonical ensemble. No CDW or pairing modulation is found in the bulk of the system, and the order parameters $\Delta_s$ and $\Delta_d$ are found to condense into the $\hat{\Delta}_{++}$ configuration.
At weak $U_{sd}$, this configuration is favored  because of an effective Josephson coupling mediated by the orbital mixing $t_{sd}$. This configuration does not produce negative geometric superfluid weight, and hence is incapable of promoting PDW order~\cite{ChenW:23}.

The charge density in the CDW phase modulates with a wavevector of $(\pi,\pi)$ and exhibits uniform amplitude in the bulk, with an anti-phase correlation between the two orbitals, where the depletion and gain of occupancy alternate between the two orbitals at every other site [Figure.~\ref{fig: other phases} (c)].
We also checked that there is no coexisting SC or spin order in this phase. 
Finally, the fully-gapped phase at zero-doping is characterized by exact one electron per-site and exponential decay in all the channels of correlation functions we examined [Fig.~\ref{fig: other phases} (b)], which potentially holds a topologically ordered phase. The exact nature of this intriguing phase requires further investigation, which is beyond the scope of the current study.  

{\bf Quantum-geometry-facilitated PDW order.---}

To explain how quantum geometry operates here, it is instructive to first recognize that geometric quantities are closely related to the interband velocity, that is, the off-diagonal elements of the velocity operator in the band-basis representation~\cite{Blount:62, Liang:17, Ahn2022NaturePhysics}:
$V^{ij}_{\mu\bk} = (\epsilon_{i\bk} -\epsilon_{j\bk}) \langle  \partial_{k_\mu} \psi_{i\bk}| \psi_{j\bk}\rangle,~(\epsilon_{i\bk} \neq \epsilon_{j\bk})$,
where $i$ labels the band index, and $|\psi_{i\bk}\rangle$ denotes the periodic part of the Bloch wavefunction on the $i$-th band, with dispersion $\epsilon_{i\bk}$. The term $i\langle \partial_{k_\mu}\psi_{i\bk}| \psi_{j\bk}\rangle$ defines a non-Abelian Berry connection between the two Bloch states. In effect, interband velocity or the corresponding non-Abelian Berry connection depicts quantum-mechanically-connected motion of electrons from different bands. 

We now illustrate the geometric effect in superconductors through one representative manifestation: Cooper pair tunneling between the Bloch bands driven by the interband velocity, as sketched in Fig.~\ref{fig:QGsketch}. This tunneling induces effective Josephson couplings between the intraband pairing order parameters $\Delta_{11}$ and $\Delta_{22}$ (where 1, 2 label the band indices), which then constitutes a unique geometric superfluid weight. Intriguingly, this Josephson coupling and hence the corresponding geometric superfluid weight changes sign by changing the relative sign of $\Delta_{11}$ and $\Delta_{22}$. If the total geometric contribution is strongly suppressed or becomes negative, states featuring spatial superconducting phase modulations, {\it i.e.}, PDW, are more favorable to form. More systematic analyses can be found in Refs.~\cite{ChenW:23, Kitamura:22, HuYJ2024}.

Due to the strong orbital mixing $t_{sd}$ in our model, the orbital-basis $\hat{\Delta}_{+-}$ pairing corresponds to a band-basis order parameter configuration where the two intraband pairings are opposite in sign. Hence, had this $\hat{\Delta}_{+-}$ pairing developed into a uniform SC in the sense that neither $\Delta_s$ nor $\Delta_d$ displays any spatial modulation, the system will receive a negative geometry-related superfluid weight. Indeed, the mean-field superfluid weight is increasingly suppressed as the pairing strength increases~\cite{ChenW:23}, thereby paving the way for PDW order to emerge. By contrast, when the repulsive pair hopping $U_{sd}$ is weak, the system tends to condense into the $\hat{\Delta}_{++}$ configuration. The resultant state does not generate negative geometric superfluid weight, and thus cannot promote PDW order.

{\bf Concluding remarks.---} We demonstrated the emergence of an intriguing $(\pi,0)$-PDW order in the strong-coupling limit of a simple two-band model on a square lattice. We established dominant PDW correlations in the phase, with $K_{sc} \sim 0.3$. To our knowledge, this PDW exhibits stronger instability than any other microscopic models previously studied using DMRG.

 The PDW obtained in DMRG arises only when quantum geometric effects would suppress the superfluid weight of the uniform SC phase. Our study therefore provides the first solid evidence supporting the quantum geometric mechanism for PDW formation. Additionally, the model studied is topologically trivial, suggesting that band topology is not a prerequisite for this intrinsic mechanism to operate. 

Looking forward, exploring the possibility of more general forms of quantum-geometry-facilitated PDW order in other multiorbital models with different lattice geometries and  interactions is a promising avenue of research. In particular, it would be interesting to design a PDW model based on the same mechanism that is free from the Quantum Monte Carlo sign problem. Moreover, it is intriguing to examine whether quantum geometry plays a pivotal role in the PDWs observed in certain cuprates and iron-based high-$T_c$ compounds~\cite{footnote1}. Finally, our study lays the groundwork for investigating whether quantum geometry can facilitate other novel states of matter, beyond PDW, in strongly interacting systems.

{\bf Acknowledgements} We acknowledge helpful discussions with Shuai Chen, Wei-Qiang Chen, Zheng-Cheng Gu, Zi-Xiang Li, Yi-Jian Hu, Wen Sun, Chang-Ming Yue and Zheng Zhu. This work is supported by NSFC under Grants No.~12374042 and No.~11904155, the Guangdong Science and Technology Department under Grant 2022A1515011948, and a Shenzhen Science and Technology Program (Grant No.~KQTD20200820113010023). Computing resources are provided by the Center for Computational Science and Engineering at Southern University of Science and Technology. The DMRG code used to simulate the two-orbital model in this work is publicly available on Github \cite{code}.

\bibliography{apssamp}

\begin{thebibliography}{91}
\bibitem [{foo()}]{footnote1}%
  \BibitemOpen
  \href@noop {} {}\bibinfo {note} {We note that the quantum geometry is expected to play an important role in iron-based superconductors, which are typically inherently multiband in nature. On the other hand, while many studies consider the cuprate superconductivity a problem of the single-band Hubbard model, quantum geometry may still feature significantly there, due to the strong hybridization of Oxygen $p$-orbitals with Cu $d$-orbital in a microscopically more accurate three-orbital (three-band) Hubbard model.}\BibitemShut {Stop}%
\end{thebibliography}

\widetext
\section{Supplemental Materials}

\setcounter{equation}{0}
\setcounter{figure}{0}
\setcounter{table}{0}
\renewcommand{\theequation}{S\arabic{equation}}
\renewcommand{\thefigure}{S\arabic{figure}}

\section{Supplementary data for Pair-density wave states}
We present extra DMRG calculation results in this section to support the robustness of the pair density wave (PDW) states.

\subsection{3-leg ladders}
The computational capacity limits our system size to $L_y \leq 4$. To carefully extrapolate the model behavior in the two-dimensional (2D) limit, we compare the numerical results from $L_y=3$ and $L_y=4$. Thus, in this file, we present some results for $L_y = 3$ under different doping levels $\delta$ and different interaction strength. Throughout this file, we set $t_s = -t_d = t_{sd} = 1$ as in the main text.

\begin{figure}[btp]
    \centering
    \includegraphics[width=0.9\textwidth]{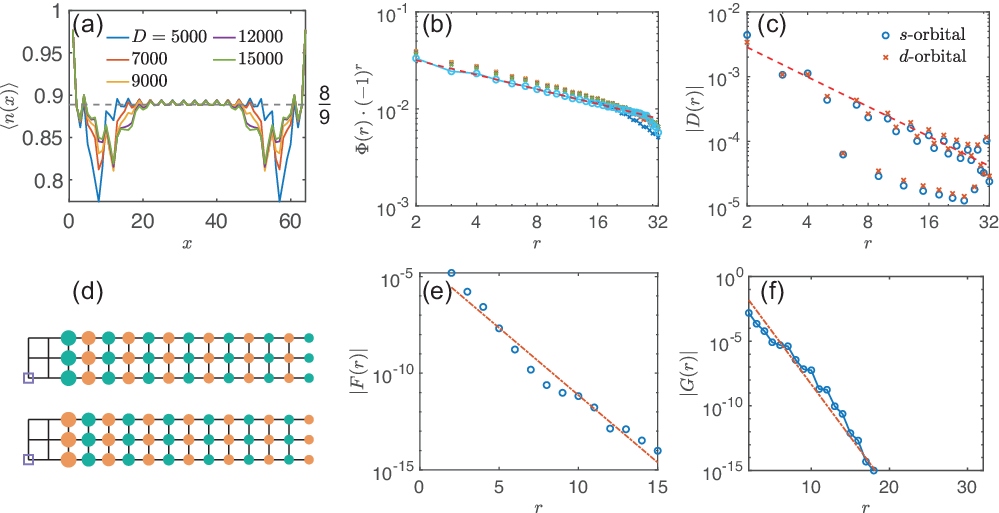}
    \caption{Numerical results of a $3 \times 64$ lattice with model parameters: $U_s = U_d = U_{sd} = 8$, and doping level $\delta = 1/8$. (a) Charge density pattern at various finite bond dimensions $D$. (b) SC correlations at finite bond dimension and extrapolation. (c) Charge density correlations. (d) Real-space SC correlations with the reference set as the $s$-orbital intraorbital pairing on site ${\bf r_0} = (16, 1)$. The area of the circle reflects the magnitude of SC correlations. The green circles indicate positive correlations, while the oranges ones indicate negative correlations. (e) and (f) show the spin and single-particle correlations. }
    \label{fig: 3-leg U8}
\end{figure}

We first present the results for a 3-leg ladder with $U_s = U_d = U_{sd} = 8$ at doping level $\delta = 1/8$, directly comparing these results with those for a 4-leg ladder with the same parameters as discussed in the main text. Figure \ref{fig: 3-leg U8} (a) shows the charge density distribution calculated using several increasingly larger sets of DMRG kept states. We note three observations from the density distributions: (i) The distributions do not converge around the boundary region with respect to the kept states but are consistent in the bulk. 
(ii) The average charge density in the bulk is approximately $8/9$, corresponding to a doping level of $\delta = 1/9$. 
(iii) There is a small charge density wave (CDW) modulation with a period of $\lambda_{\mathrm{CDW}} = 3$. The doping level $\delta = 1/9$ in the bulk can also be confirmed by the prior belief that the stripe filling is $1/L_y$ in the PDW phase, which has been observed in other cases. These three observations suggest that under this set of parameters, the model on a 3-leg ladder ($L_y = 3$) tends to form phase separation. We also obtain that the following correlations measured in the bulk of the system reflect the behavior at a doping level $\delta = 1/9$.

We measured the SC, charge, spin, and single-particle correlations, as shown in Fig. \ref{fig: 3-leg U8}(b-f). The behavior of the correlation functions is consistent with that in $L_y = 4$ systems. We again observe a $(\pi, 0)$-PDW with a power-law decay in the SC correlation. The charge mode is gapless, characterized by a power-law decay in the charge correlation. The spin and single-particle degrees of freedom are all gapped and characterized by exponential decay correlations. The SC and density correlations give Luttinger parameters $K_{sc} = 0.50$ and $K_c = 1.52$, respectively. Here, the tails of the SC and density correlations slightly deviate from the dashed guideline due to the phase separation.

From the above calculations, we conclude that a strong PDW phase appears in the $L_y = 3$ system under these parameters. The Luttinger parameter $K_{sc}$ is larger than that in the $L_y = 4$ system discussed in the main text with doping levels $\delta = 1/8$ and $1/32$, suggesting that the SC becomes stronger as $L_y$ increases, thus further indicating the presence of long-range SC order in the 2D limit.

\begin{figure}[btp]
    \centering
    \includegraphics[width=0.9\textwidth]{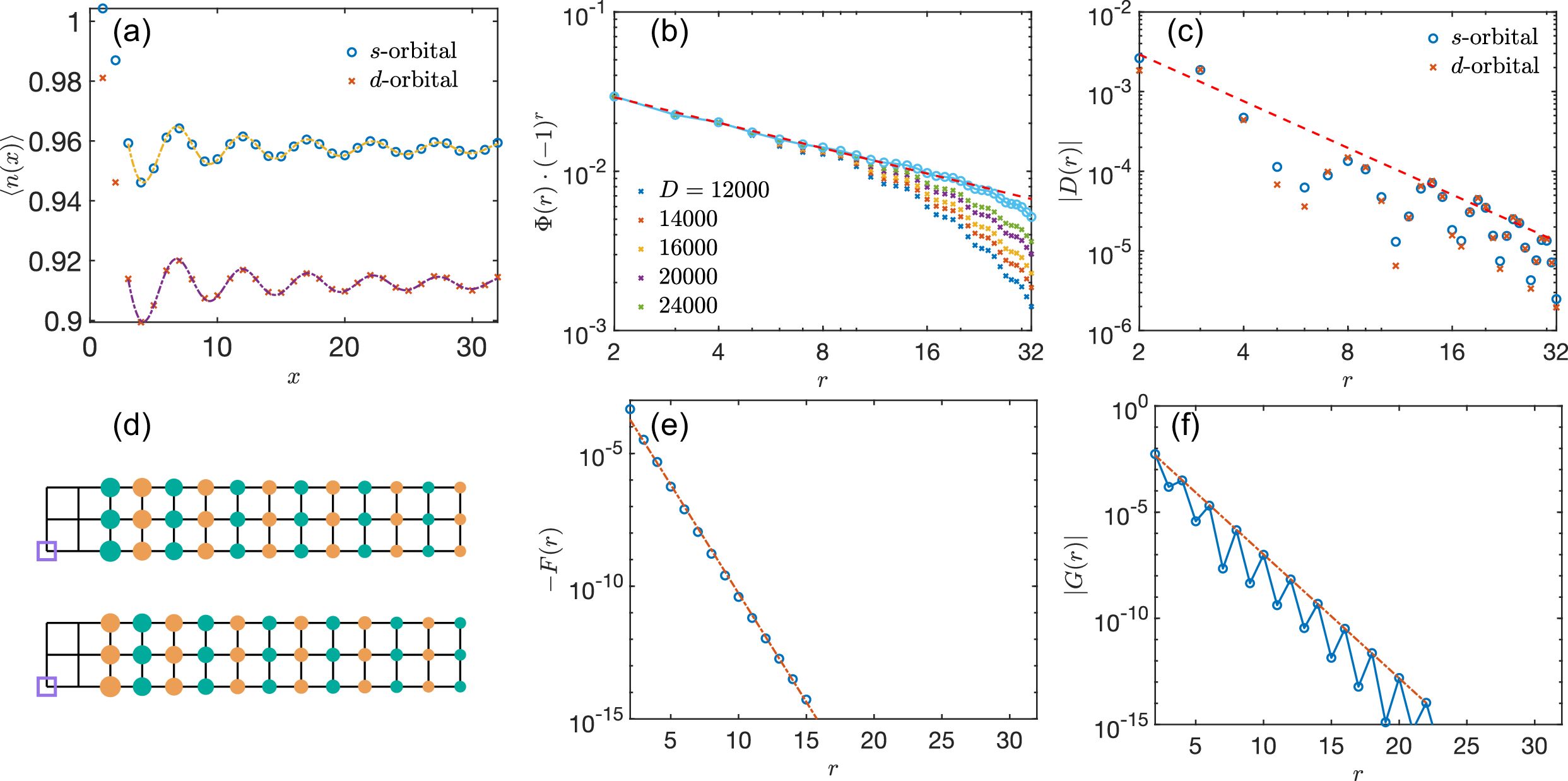}
    \caption{Correlations on a $3 \times 64$ lattice. Model parameters: $U_s = 3.8$, $U_d = 3.7$, $U_{sd} = 4.0$, and doping level $\delta = 1/16$. (a) Charge density profile and Friedel oscillation. (b) SC correlations. (c) Charge density correlations. (d) SC correlations a 2D top-view representation. (e) Spin correlation. (f) Single-particle correlation.}
    \label{fig: 3leg U4}
\end{figure}

To further explore regions with relatively weaker interactions, we examine the calculation results for $t_s = -t_d = t_{sd} = 1$ and $U_s = 3.8$, $U_d = 3.7$, $U_{sd} = 4.0$, as shown in Figure \ref{fig: 3leg U4}. The mean-field solution for these parameters has been considered in Ref.~[41] and indicates a $(\pi,0)$-PDW order in the ground state. The results here are qualitatively similar to the PDW state investigated in the strong interaction regime. 

The charge density wave (CDW) shows clear Friedel oscillations and decaying CDW amplitudes as one moves into the bulk of the system, with a modulation periodicity $\lambda_{\mathrm{CDW}} = \frac{16}{3}$, giving a stripe filling of $\frac{1}{L_y}$. The quasi-long-range $(\pi,0)$-PDW and density correlations provide Luttinger parameters $K_{sc} =0.53$ and $K_c = 1.94$, respectively. The $K_c$ fitted from the density correlation is consistent with that fitted from the Friedel oscillation up to two decimal places. Given the fact that both spin and single-particle modes are gapped, the product $K_{sc}K_c \approx 1 $ strongly indicates a Luther-Emery liquid state in this regime.

\subsection{Grand canonical ensemble DMRG simulation}

\begin{figure}[htbp]
    \centering
    \includegraphics[width=0.9\textwidth]{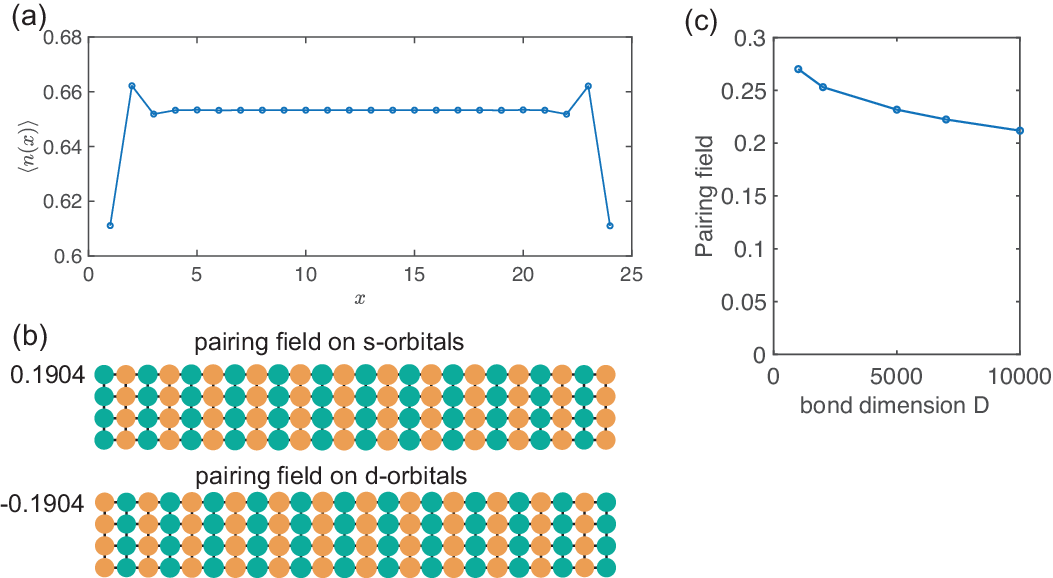}
    \caption{Charge density and SC pairing field calculated in the grand canonical ensemble. Model parameters: $U_s = U_d = 8$, $U_{sd} = 4$, $\mu_s = \mu_d = -6$, corresponding to a doping level $\delta \approx 0.35$. (a) Charge density pattern for kept state $D=10000$. (b) SC pairing field. (c) The magnitude of SC pairing field vs. DMRG kept state $D$.}
    \label{fig: PDW grand canonical ensemble}
\end{figure}

The existence of the PDW phase is further confirmed by the grand canonical ensemble DMRG simulation. To adjust the doping level, we add a chemical potential term $-\mu_s \hat{N}_s - \mu_d \hat{N}_d$ to the Hamiltonian, where $\hat{N}_s$ and $\hat{N}_d$ represent the total particle numbers in the $s$- and $d$-orbitals, respectively. We demonstrate a representative point in the phase diagram with model parameters $U_s = U_d = 8$, $U_{sd} = 4$, and $\mu_s = \mu_d = -6$, as shown in Figure~\ref{fig: PDW grand canonical ensemble}. The doping level for this set of parameters is approximately 0.35. 

The charge density distribution reveals no CDW in the ground state [Figure~\ref{fig: PDW grand canonical ensemble} (a)]. The on-site pairing field $\langle c^\dagger_{i, \uparrow, \alpha} c^\dagger_{i, \downarrow, \alpha} \rangle$, shown in Figure~\ref{fig: PDW grand canonical ensemble} (b), exhibits a $(\pi,0)$-PDW modulation, consistent with the PDW correlation measured in the canonical ensemble. The pairing fields between different layers exhibit a phase difference, forming the $\hat{\Delta}_{+-}$ configuration, which facilitates the formation of the PDW. Additionally, we plot the dependence of the on-site pairing field strength on the bond dimension $D$ in Figure~\ref{fig: PDW grand canonical ensemble} (c). The "shoulder" strongly indicates the leading instability of the SC order [28, 85].

\section{Establishment of Uniform Superconductivity in the Small-$U_{sd}$ Region}

Due to space limitations, we previously did not provide sufficient data to support the existence of uniform superconductivity in the small $U_{sd}$ region. In this section, we present representative results illustrating the uniform superconducting (SC) states in the phase diagram. All data presented here are gathered from $L_y = 4$ systems.

Before presenting the results, we discuss a technical detail. As the system transitions from the PDW phase to the uniform phase by decreasing $U_{sd}$, the truncation error in DMRG calculations abruptly increases from the order of $10^{-6}$ to $10^{-4}$, indicating a phase transition nearby. However, this increase in truncation error also makes the calculations more difficult to converge. With truncation errors of this magnitude, the extrapolation of the correlation function may not be reliable. Therefore, we need to carefully analyze the raw data of the correlations.

\begin{figure}[tbp]
    \centering
    \includegraphics[width=1.0\textwidth]{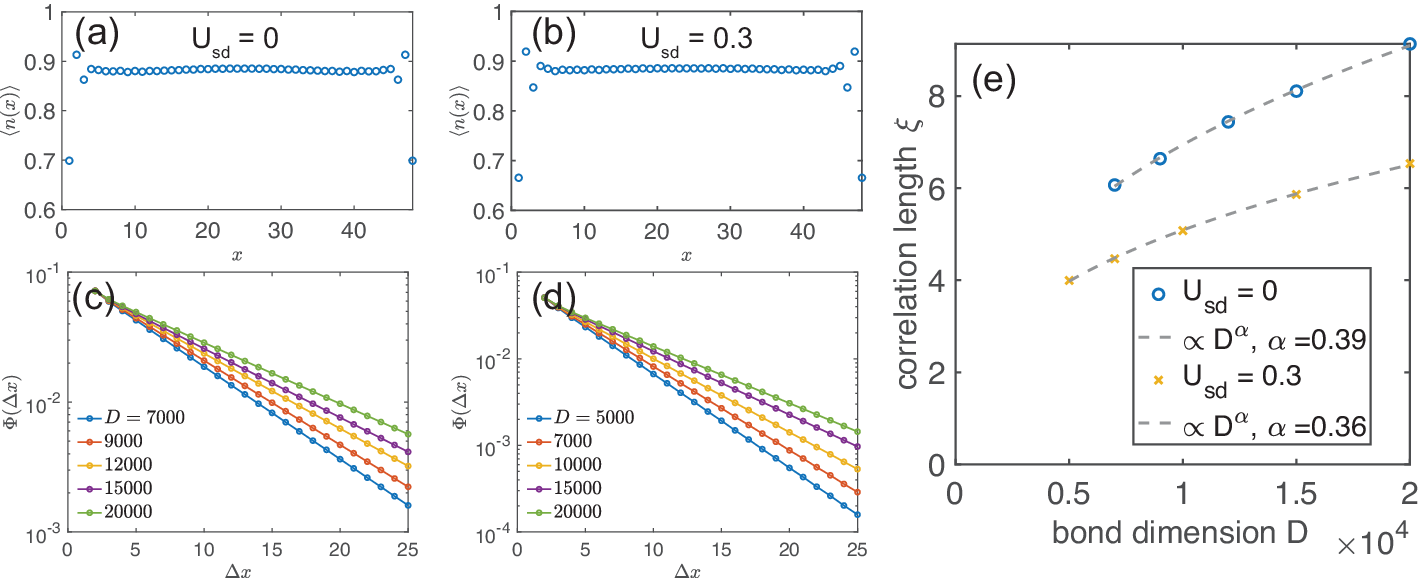}
    \caption{Charge density pattern and SC correlation calculated in the uniform SC phase. The model parameters are chosen from the phase diagram in Figure 2 in the main text, with doping level $\delta = 1/8$ and $U_{sd}=0$ or $0.3$. (a,b) show the charge density pattern. (c,d) show the SC correlation in the $s$-orbital obtained by a series of gradually increasing DMRG bond dimensions. (e) shows the bond dimension dependence of the SC correlation length $\xi$.}
    \label{fig: uniform SC canonical}
\end{figure}

We show the charge and SC properties of the uniform SC states in Fig.~\ref{fig: uniform SC canonical}. We choose model parameters $\delta = 1/8$ and $U_{sd} = 0$ or $0.3$ from the phase diagram in Figure 2. The charge density exhibits no apparent modulation for both parameters, indicating the absence of CDW order. While the SC correlations decay exponentially under finite bond dimensions, they visibly enhance with increasing bond dimension. We fit the SC correlation lengths and plot their dependence on the bond dimensions in Fig.~\ref{fig: uniform SC canonical}(e). The dependence of the correlation length on the bond dimensions can be well fitted by $\xi \propto D^\alpha$, with $\alpha = 0.39$ and $0.36$ for the respective cases. This suggests that the SC correlation may diverge at the infinite bond dimension. Assuming that the SC correlations decay in a power-law manner at infinite bond dimension, a brute-force power-law fitting of the correlations at $D=20000$ yields an upper bound estimation for the Luttinger parameter $K_{sc} \leq 0.74$ and $K_{sc} \leq 1.21$.

\begin{figure}[htbp]
    \centering
    \includegraphics[width=0.4\textwidth]{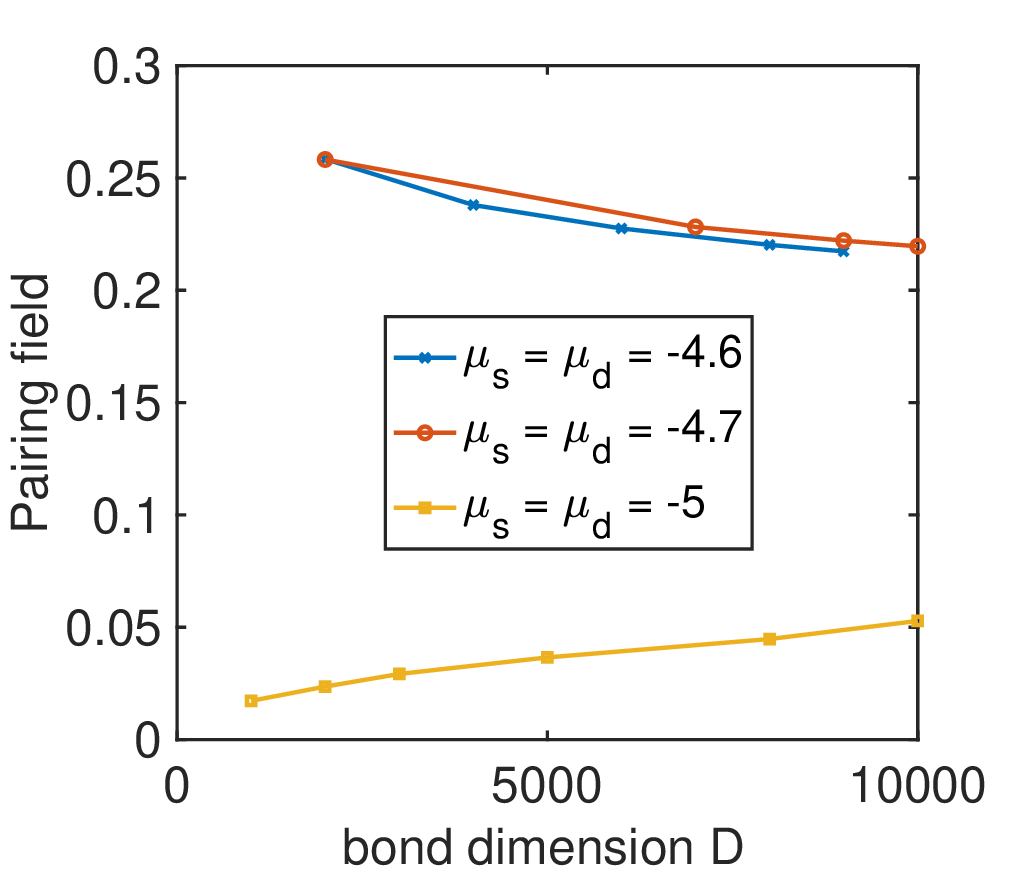}
    \caption{The magnitude of uniform SC pairing field $v.s.$ DMRG bond dimension $D$. $U_s = U_d = 8$, and $U_{sd} = 0.3$. }
    \label{fig: uniform pairing field grand canonical ensemble}
\end{figure}

The existence of uniform superconductivity (SC) has also been confirmed by grand canonical ensemble calculations, as shown in Fig. 5(a) in the main text. Here, we present the corresponding bond dimension dependence of the pairing field strength in Fig.~\ref{fig: uniform pairing field grand canonical ensemble}. In the figure, we include two additional sets of calculations under different chemical potentials. For the cases with $\mu_s = \mu_d = -4.6$ or $-4.7$, the pairing field strength exhibits a "shoulder" behavior [28, 85]. For the case with $\mu_s = \mu_d = -5$, the pairing field strength increases as the bond dimension $D$ increases. All these cases indicate the leading instability toward uniform SC order.

\section{Data extrapolations in DMRG}
We use a quadratic function to extrapolate the expectation values in DMRG calculations. We denote $ A(\epsilon) $ as the truncation error dependence of the measurement results for the observable $\hat{A}$ with successively changing bond dimension $D$. The extrapolated measurement $ A(0) $, corresponding to the infinite bond dimension, is obtained by quadratic fitting as follows:

\begin{equation}
A(\epsilon) = A(0) + b\epsilon + c\epsilon^2.
\end{equation}

\end{document}